\RequirePackage{fix-cm}
\documentclass[twocolumn,epjc3]{svjour3}  
\smartqed  
\RequirePackage{graphicx}
\usepackage{units}
\usepackage{siunitx}
\usepackage{bm}
\usepackage{booktabs}
\usepackage{subfig}
\usepackage{textcomp}
\usepackage{multirow}
\usepackage{placeins}

\usepackage[version=4]{mhchem}
 \RequirePackage{mathptmx}      
\journalname{Eur. Phys. J. C}
\begin{document}

\title{Potential for a precision measurement of solar $pp$ neutrinos in the Serappis Experiment
}


\author{
Lukas Bieger\thanksref{54}
\and
Thilo Birkenfeld\thanksref{48}
\and
David Blum\thanksref{54}
\and
Wilfried Depnering\thanksref{52}
\and
Timo Enqvist\thanksref{42}
\and
Heike Enzmann\thanksref{52}
\and
Feng Gao\thanksref{48}
\and
Christoph Genster\thanksref{50}
\and
Alexandre G\"{o}ttel\thanksref{50,48}
\and
Christian Grewing\thanksref{51}
\and
Maxim Gromov\thanksref{67}
\and
Paul Hackspacher\thanksref{52}
\and
Caren Hagner\thanksref{49}
\and
Tobias Heinz\thanksref{54}
\and
Philipp Kampmann\thanksref{50}
\and
Michael Karagounis\thanksref{51}
\and
Andre Kruth\thanksref{51}
\and
Pasi Kuusiniemi\thanksref{42}
\and
Tobias Lachenmaier\thanksref{54}
\and
Daniel Liebau\thanksref{51}
\and
Runxuan Liu\thanksref{50,48}
\and
Kai Loo\thanksref{52}
\and
Livia Ludhova\thanksref{50,48}
\and
David Meyh\"{o}fer\thanksref{49}
\and
Axel M\"{u}ller\thanksref{54}
\and
Pavithra Muralidharan\thanksref{51}
\and
Lothar Oberauer\thanksref{53}
\and
Rainer Othegraven\thanksref{52}
\and
Nina Parkalian\thanksref{51}
\and
Yatian Pei\thanksref{10}
\and
Oliver Pilarczyk\thanksref{52}
\and
Henning Rebber\thanksref{49}
\and
Markus Robens\thanksref{51}
\and
Christian Roth\thanksref{51}
\and
Julia Sawatzki\thanksref{53}
\and
Konstantin Schweizer\thanksref{53}
\and
Giulio Settanta\thanksref{50}
\and
Maciej Slupecki\thanksref{42}
\and
Oleg Smirnov\thanksref{67}
\and
Achim Stahl\thanksref{48}
\and
Hans Steiger\thanksref{52,53}
\and
Jochen Steinmann\thanksref{48}
\and
Tobias Sterr\thanksref{54}
\and
Matthias Raphael Stock\thanksref{53}
\and
Jian Tang\thanksref{20}
\and
Eric Theisen\thanksref{52}
\and
Alexander Tietzsch\thanksref{54}
\and
Wladyslaw Trzaska\thanksref{42}
\and
Johannes van den Boom\thanksref{51}
\and
Stefan van Waasen\thanksref{51}
\and
Cornelius Vollbrecht\thanksref{50,48}
\and
Christopher Wiebusch\thanksref{48}
\and
Bjoern Wonsak\thanksref{49}
\and
Michael Wurm\thanksref{52}
\and
Christian Wysotzki\thanksref{48}
\and
Yu Xu\thanksref{50,48}
\and
Ugur Yegin\thanksref{51}
\and
Andre Zambanini\thanksref{51}
\and
Jan Z\"ufle\thanksref{54}
}

\institute{
Institute of High Energy Physics, Beijing, China\label{10}
\and
Sun Yat-Sen University, Guangzhou, China\label{20}
\and
University of Jyvaskyla, Department of Physics, Jyvaskyla, Finland\label{42}
\and
III. Physikalisches Institut B, RWTH Aachen University, Aachen, Germany\label{48}
\and
Institute of Experimental Physics, University of Hamburg, Hamburg, Germany\label{49}
\and
Nuclear Physics Institute IKP-2, Forschungszentrum J\"{u}lich GmbH,  J\"{u}lich, Germany\label{50}
\and
Central Institute of Engineering, Electronics and Analytics - Electronic Systems(ZEA-2), Forschungszentrum J\"{u}lich GmbH, J\"{u}lich, Germany\label{51}
\and
Institute of Physics and Excellence Cluster PRISMA$^+$, Johannes-Gutenberg Universität Mainz, Mainz, Germany\label{52}
\and
Technische Universit\"{a}t M\"{u}nchen, M\"{u}nchen, Germany\label{53}
\and
Eberhard Karls Universit\"{a}t T\"{u}bingen, Physikalisches Institut, T\"{u}bingen, Germany\label{54}
\and
Joint Institute for Nuclear Research, Dubna, Russia\label{67}
}







\date{Received: date / Accepted: date}

\maketitle

\begin{abstract}
The Serappis (SEarch for RAre PP-neutrinos In Scintillator) project aims at a precision measurement of the flux of solar $pp$ neutrinos on the few-percent level. Such a measurement will be a relevant contribution to the study of solar neutrino oscillation parameters and a sensitive test of the equilibrium between solar energy output in neutrinos and electromagnetic radiation (solar luminosity constraint). The concept of Serappis relies on a small organic liquid scintillator detector ($\sim$20\,m$^3$) with excellent energy resolution ($\sim$2.5\% at 1\,MeV), low internal background and sufficient shielding from surrounding radioactivity. This can be achieved by a minor upgrade of the OSIRIS facility at the site of the JUNO neutrino experiment in southern China. To go substantially beyond current accuracy levels for the $pp$ flux, an organic scintillator with ultra-low \ce{^{14}C} levels (below $10^{-18}$) is required. The existing OSIRIS detector and JUNO infrastructure will be instrumental in identifying suitable scintillator materials, offering a unique chance for a low-budget high-precision measurement of a fundamental property of our Sun that will be otherwise hard to access.

\end{abstract}

\section{Introduction}
\label{sec:intro}
The flux of solar $pp$ neutrinos emitted during the main fusion process of two protons to a deuteron ({\it pp-I}) is probably the most basic prediction of the Standard Solar Model (SSM). Not only is the nuclear physics involved well understood but more importantly the predicted $pp$ neutrino flux is closely coupled to the solar luminosity constraint that can be obtained from the measurement of the Sun's electromagnetic emissions. Consequently, the SSM cites an uncertainty of $\pm$0.6\% for the $pp$ neutrino flux prediction (e.g.~\cite{Serenelli:2016dgz}).

Experimental constraints are by comparison substantially weaker. Only a handful of experiments have been able to perform measurements of solar neutrinos that included the $pp$ component. The radiochemical experiments Gallex/GNO and SAGE that employed \ce{^{71}Ga} as solar neutrino target were able to measure the total rate of solar neutrinos with energies above 233 keV. Since this was dominated by $pp$ neutrino interactions, this provided a constraint on the $pp$ flux at $\sim$20\% accuracy \cite{Kirsten:2008zz}. In 2014, the Borexino collaboration published the first real-time measurement of the $pp$ component to $\sim$10\% \cite{Bellini:2014uqa}, further refining the result in a follow-up publication with only a mild gain in uncertainty levels \cite{Agostini:2018uly}. While some improvement cannot be excluded from the yet-to-be-released analysis of the full Borexino data set, the measurement uncertainties can be expected to be still far greater than the SSM prediction.

That makes a precision measurement of the $pp$ neutrino rate on Earth a most interesting endeavor. On the first level, a more accurate knowledge of the $pp$ electron neutrino ($\nu_e$) flux can be used to constrain the solar mixing angle $\theta_{12}$ at high precision \cite{Bahcall:2004mz}. Due to their low energy ($Q_{pp}<420$\,keV), $pp$ neutrinos provide a clean measurement of the survival probability $P_{ee}$ for vacuum oscillations. Since the current uncertainty level on the oscillation amplitude $\sin^2\theta_{12}$ is about $4\%$, a similar level of measuring accuracy will be required to improve on the current global uncertainties \cite{Esteban:2020cvm}.

With the upcoming start of operation of the JUNO experiment, a true precision result of about 0.5\% on the solar oscillation amplitude is likely only a few years away \cite{2022103927}. Since JUNO bases its measurements on reactor $\bar\nu_e$ oscillations, a comparison to solar neutrino results will still be useful to test neutrino-antineutrino symmetry in oscillations. More interestingly, JUNO's result on $P_{ee}$ can be employed to correct the solar $pp$ flux measurement and compare the result to the fairly model-independent SSM prediction. If a divergence was found, it would be a sign of ''invisible'' energy losses from the Sun, e.g.~mediated by thermal production of axions or hidden photons \cite{Vinyoles:2015aba}. A measurement on the level of 1\% would enable to set rather tight constraints on these or comparable effects.

How to improve on current experimental limits? The result from Borexino is to a large degree limited by the measurement systematics caused by a large intrinsic background source at $pp$ neutrino energies: the $\beta$-decay of \ce{^{14}C} with an endpoint of 156\,keV. While the radioactive isotope is only present at the level of $2.5\times10^{-18}$ compared to the stable \ce{^{12}C}, the activity in the entire liquid scintillator target of Borexino amounts to $\sim$100\,Bq, a formidable rate when compared to the several hundred events per day expected from the elastic scattering of $pp$ neutrinos off the target electrons. However, the abundance of \ce{^{14}C} in organic scintillator depends on a number of factors (Sec.~\ref{sec:c14}) and can be expected to be substantially lower in case a suitable source of crude oil is found.

The present paper investigates the possibility to turn a 20-ton liquid scintillator detector into a competitive $pp$ neutrino experiment. Apart from low-\ce{^{14}C} scintillator, such a detector will profit from ample shielding of external gamma-rays and high photoactive coverage to achieve excellent photon collection and thus energy resolution at the low $pp$ energies. Maybe somewhat counter-intuitively, it also will benefit from its relatively little target size that will greatly reduce the frequency and thus importance of accidental coincidences of \ce{^{14}C} decays as a relevant contribution to the low-energy spectrum. Such an experiment, dubbed Serappis for {\it SEarch for RAre PP neutrinos In Scintillator}, could be realized as an upgrade of the OSIRIS pre-detector of the JUNO experiment. OSIRIS' primary purpose is the investigation of the radiopurity of the JUNO liquid scintillator \cite{2022103927,JUNO:2021wzm}. As will be explained in more detail in Section~\ref{sec:setup}, there are good reasons to assume that a moderate upgrade of OSIRIS to Serappis will be sufficient to outperform the much larger JUNO detector in the specific case of a $pp$ neutrino measurement.

At present, the most serious competitor for a high-preci-sion $pp$ measurement is the planned DARWIN dark matter experiment \cite{Aalbers:2020gsn}. The 50-ton liquid xenon TPC features a low-energy detection window for $pp$-$\nu$ induced electron recoils in the energy range below 200\,keV. While the statistical uncertainty after several years of measuring is expected to be very low (sub-percent level), this measurement at the low end of the $pp$ spectrum features a strong degeneracy between the flux normalization and the dependence of the differential cross section for $\nu_ee$-scattering on the exact value of the Weinberg angle $\theta_W$. The corresponding effect on the spectrum of the recoil electrons and hence $pp$ event rate extracted from a spectral fit in a Serappis measurement will be discussed in Section \ref{sec:weinberg}.  

The paper is structured as follows: In Section \ref{sec:setup}, we describe the envisaged detector layout of Serappis and point out in which respects it is advantageous for a $pp$ neutrino detection in comparison to Borexino and JUNO. The prospects of identifying the source for a low-\ce{^{14}C} scintillator are discussed in Section \ref{sec:c14}. The detector simulation and production of event samples is described in Section \ref{sec:sim}. Section \ref{sec:sensitivity} investigates the expected sensitivity as a function of \ce{^{14}C} abundance, energy resolution, measuring time and other important experimental parameters. We conclude in Section \ref{sec:conclusions}.

\section{Experimental Layout}
\label{sec:setup}
The Serappis experiment will profit from the existing infrastructure of the JUNO reactor neutrino experiment \cite{2022103927,An:2015jdp}. JUNO will feature an elaborate liquid scintillator (LS) system for mixing and purifying the basic organic components, i.e.~linear alkylbenzene (LAB) with the fluor PPO (g/l) and the wavelength-shifter Bis-MSB (mg/l) as additives. Before filling the product LS into the JUNO Central Detector, it will be monitored for its radiopurity in the 20-m$^3$ pre-detector OSIRIS to make sure it meets JUNO's physics requirements. In this section, we first describe the current layout of the OSIRIS detector (Sec.~\ref{sec:osiris}). A series of moderate upgrades planned for the detector hardware (described in Sec.~\ref{sec:serappis}) will improve both the detector performance and its shielding from external backgrounds, constituting our default Serappis setup. It is important to note that the availability of the JUNO LS systems is crucial both for identifying a low-\ce{^{14}C} scintillator (Sec.~\ref{sec:c14}) and to remove other radioactive backgrounds (e.g.~U/Th chain elements) from the LS (Secs.~\ref{sec:internal} and \ref{sec:sensitivity}).   

\subsection{OSIRIS Design}
\label{sec:osiris}

\begin{figure}[t!]
    \centering
    \includegraphics[width=0.48\textwidth]{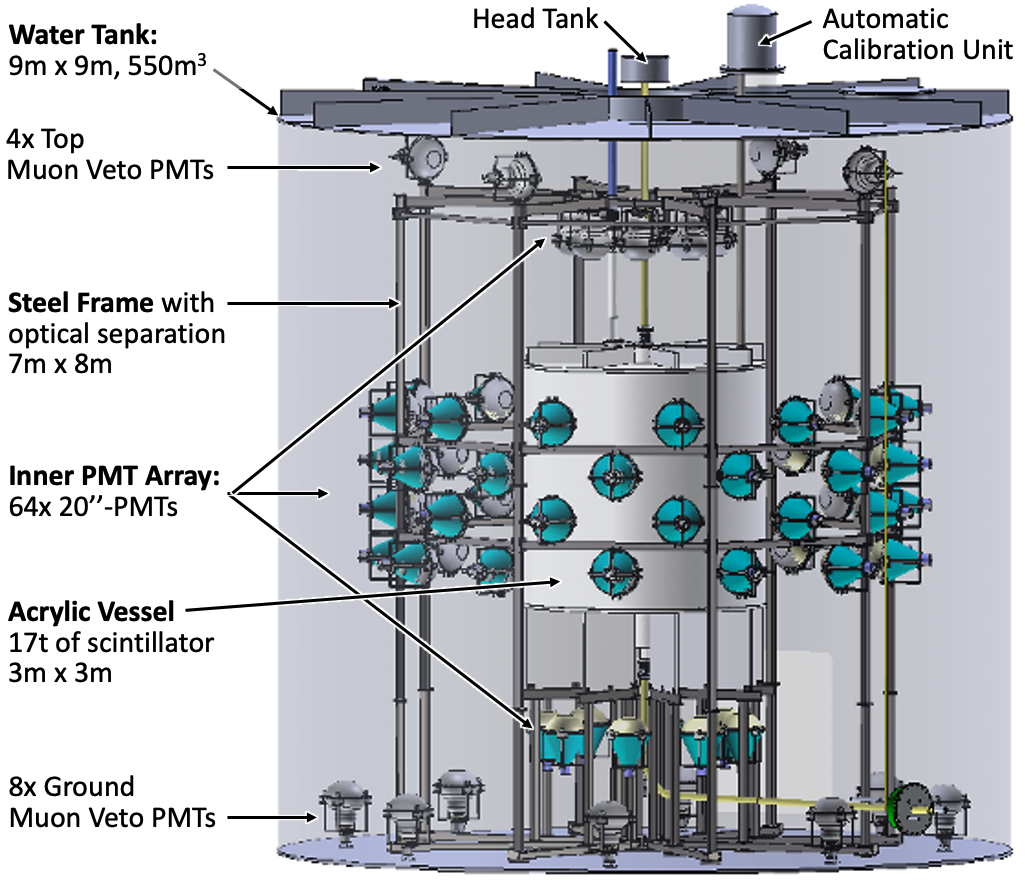}
    \caption{Layout of the OSIRIS setup.}
    \label{fig:osiris_layout}
\end{figure}

The OSIRIS detector will be located in the underground JUNO Liquid Scintillator Hall that provides 693\,m of rock overburden (1800\,m.w.e.) \cite{2022103927}. A sketch of the experimental layout is shown in Fig.~\ref{fig:osiris_layout}. The three main mechanical components of the OSIRIS experiment are: a cylindrical Acrylic Vessel (AV) of 3\,m height and diameter each that holds 18 tons of liquid scintillator; a surrounding cylindrical Water Tank (WT) of 9\,m by 9\,m that holds a 3-m wide ultrapure water buffer for shielding of external gamma rays; and a stainless Steel Frame (SF) holding the Photomultiplier Tubes (PMTs) and other auxiliary systems. OSIRIS is equipped with a total of 76 dynode PMTs (R12860) with 20'' diameter and high quantum efficiency ($\sim$28\%). The inner array for detection of the scintillation light is formed by 64 PMTs arranged around the AV at 1.3\,m distance from its surface.  Twelve PMTs are positioned on the floor and below the lid of the WT and use the outer layer of ultrapure water as a Cherenkov radiator to veto incident cosmic muons. The two subdetectors are optically separated by black-and-white PET sheets spanned between the profiles of the Steel Frame. 


\begin{figure}[t!]
    \centering
    \includegraphics[width=0.48\textwidth]{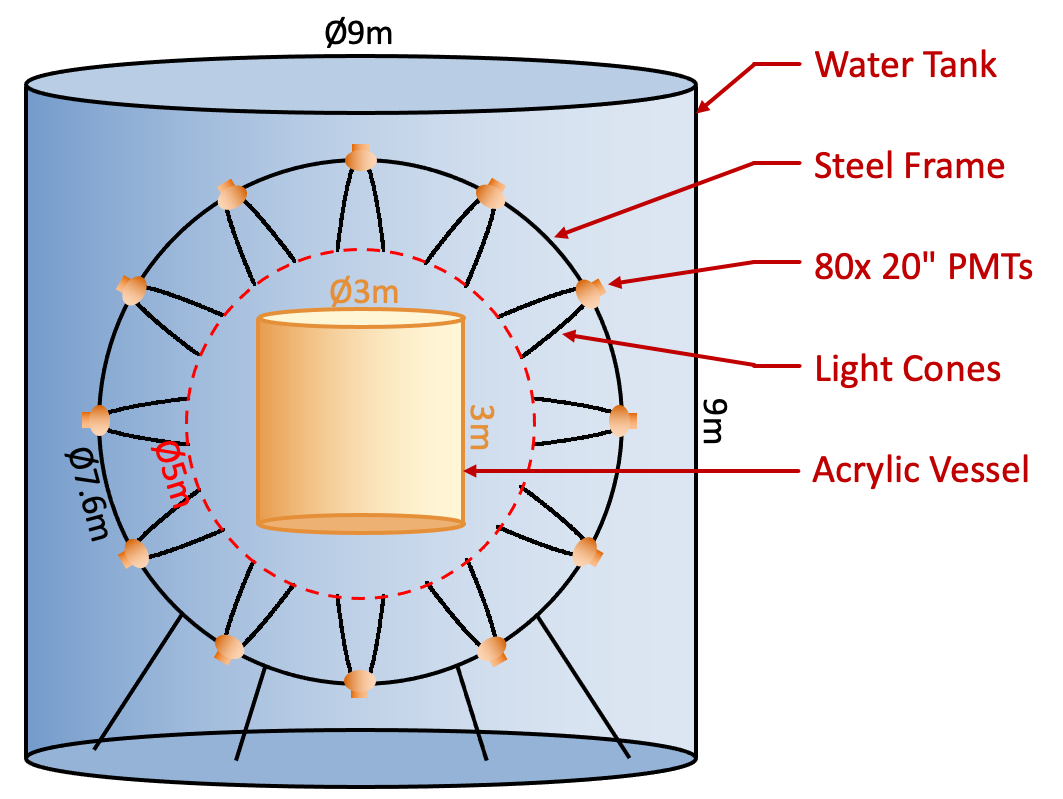}
    \caption{Conceptual layout of the Serappis upgrade. PMTs are moved further from the Acrylic Vessel and equipped with light cones.}
    \label{fig:serappis_layout}
\end{figure}

The most important performance parameters are summarized in Table ~\ref{tab:improvements}. With $\sim$9\% photo coverage, the inner PMT array provides a photoelectron (p.e.) yield of about 275\,p.e./MeV.
The signals of the ''intelligent'' PMTs are digitized directly at their base and transferred to the outside electronics via ethernet. An online software trigger selects events above a threshold of 5\ p.e. (20\,keV) coincident within 70\,ns. Full pulse shapes with a sampling rate of 500\,MS/s are written to disc. Based on the hit information acquired, we extrapolate an energy resolution of 6\% and a spatial resolution of 14\,cm (1D) at a visible energy of 1\,MeV (electron equivalent) \cite{Li:2021oos}. The water buffer reduces the background rates induced primarily by the surrounding rock and PMT glass to a level of 6\,Bq inside the LS volume. A more detailed description of the detector layout and the expected performance of the OSIRIS detector can be found in Ref.~\cite{JUNO:2021wzm}. 

\subsection{Upgrade to the Serappis experiment}
\label{sec:serappis}

The main obstacles to a successful detection of $pp$ neutrinos in the OSIRIS setup just described are the high external gamma background level as well as the relatively poor energy resolution. However, the situation can be substantially improved by a series of minor modifications to the setup.

A conceptual drawing of the improved Serappis setup is shown in Fig.~\ref{fig:serappis_layout}. The main sources of external background are gamma-rays emitted by the U/Th and \ce{^{40}K}-rich PMT glass and the cavern rock surrounding the detector. A reduction by 3-4 orders of magnitude in the external background is required in order to bring it to a level comparable to the solar neutrino signal. This can be achieved by additional external shielding corresponding to about $\sim$1.2\,m water equivalent, that could e.g.~be realized by a combination of an outer layer of low-radioactivity concrete blocks ($\sim$40\,cm) supported by an inner layer of steel shielding. Note that the additional shielding from below has been already provided by an array of steel plates inserted in the concrete floor below the detector. Moreover, shielding from PMT radioactivity can be improved by re-arranging the inner PMT array,~i.e.~mounting them to a new Steel Frame providing a spherical surface at 3.8\,m distance from the detector center.  

The reduced solid angle coverage of the PMTs would in principle translate to a lower light collection efficiency. However, the larger distance to the AV surface opens as well the opportunity to employ reflective light concentrators (''Winston cones'') in front of the PMTs. Those have been used by a number of neutrino detectors, including Borexino and its Counting Test Facility (CTF), where light cones provided a factor 2.5 and 8.8 in increased  collection area, respectively \cite{OBERAUER2004453}. For the dimensions of the Serappis setup, the optimum concentrators would be about 1\,m in front opening diameter and 1.3\,m in length. The corresponding enlargement of the photocollective area by a factor 4.3 as well as the effectively smaller solid angle would correspond to $\sim$75\% photoactive coverage. The maximum viewing angle of the cones of $\sim$30$^\circ$ will still be sufficient to observe the aspired fiducial volume (Sec.~\ref{sec:sensitivity}) for $pp$ neutrino detection.

The positive impact of these modifications is summarized in the last column of Table ~\ref{tab:improvements}. Neglecting reflection losses (they could be $\sim 10\%$ \cite{OBERAUER2004453}), the effective p.e.~yield might reach as high as 2,000 p.e./MeV, corresponding to substantial improvements in detector resolution: regarding for the moment only the influence of photon statistics, an energy resolution of 2.2\% and a vertex resolution of 5\,cm (both at 1\ MeV) can be expected for the Serappis detector, while the gamma background would be reduced to less than 200 counts per day in the entire LS volume. The impact of these improved performance parameters on $pp$ neutrino detection will be substantial and is investigated in detail in Sec.~\ref{sec:variation}.

\begin{table}[t]
    \centering
    \caption{Expected performance parameters of the upcoming OSIRIS and the upgraded Serappis detector}
    \label{tab:improvements}
    \begin{tabular}{l|ccl}
     \bf Detector Property    & \bf OSIRIS & \bf Serappis & \bf Unit\\
         \hline
     Effective photo coverage   & 9\%  & $\leq$75\%    \\
     Photo electron yield $Y_{pe}$ & 275  & $\leq$2,000 & p.e./MeV\\
     Energy resolution ($\sigma$) & 6\% & $\sim$2.2\% & at 1\,MeV\\
     Position resolution ($\sigma$)& 13\,cm & $\sim$5\,cm &  at 1\,MeV\\
     External background rate    & 6\,Bq & $\leq$2.3\,mBq &\\
    \end{tabular}
\end{table}

\subsection{Comparative performance}

To appreciate why a small detector like Serappis ($\sim20$ ton) might actually achieve a better sensitivity for $pp$ neutrinos than the larger Borexino ($\sim278$ ton) or giant JUNO experiments ($\sim20$ kton), it is important to regard both detector performance and background conditions in the respective experiments.

In the case of Borexino, the precision of the $pp$ rate results is on the 10\% level. While a further increase in exposure would likely permit to improve the result by a few percent, sensitivity will be lastly limited by two factors: the p.e.~yield, starting at about 500\ p.e./MeV and slowly deteriorating over time due to the loss of functioning PMTs, governs the resolution of the transition from the \ce{^{14}C} to the $pp$ spectrum and thus the sensitivity of the fit, as exemplified by Table~\ref{tab:delta_pp}(b). And crucially, the pile-up of \ce{^{14}C} decays with itself and other spectral components that has to be taken into account in the Borexino event spectrum and causes additional systematic uncertainty \cite{Bellini:2014uqa}.

From the point of view of the energy resolution, JUNO in principle starts from a better position since a p.e.~yield on the level of 1,345\ pe/MeV is expected \cite{2022103927}. Despite this excellent photon statistics, the high dark rate of the 17,612 PMTs (especially the NNVT type) is expected to cause a substantial decrease in effective energy resolution at these lowest energies \cite{2022103927}. Moreover, due to the huge volume and resulting \ce{^{14}C} background rate, pile-up at low energies will be happening at even higher frequency than in Borexino.

In summary, the combination of a small detection volume with a very high p.e.~collection efficiency makes the Serappis concept potentially superior for the specific case of solar $pp$ neutrino detection. Moreover, the relatively small LS mass required opens the possibility to very carefully pre-select the organic materials of which the LS is made. As outlined in the following section, this may permit to substantially reduce the most important background to $pp$ detection, i.e., \ce{^{14}C}.

\section{Low \ce{^{14}C} Scintillator}
\label{sec:c14}

The sensitivity to $pp$ neutrinos in organic scintillator could be substantially enhanced by a reduction of the background level generated by intrinsic \ce{^{14}C} decays \cite{Smirnov:2003by,Derbin:2004ma}. By everyday standards, the levels of \ce{^{14}C} compared to \ce{^{12}C} that have been found in the large LS neutrino detectors are extremely low, reaching from $R(\ce{^{14}C}:\ce{^{12}C})=R\approx1.9\times10^{-18}$ in case of the CTF over $R\approx 2.7\times10^{-18}$ from Borexino to $R\approx4\times10^{-18}$ observed in KamLAND \cite{Alimonti:1998rc}\cite{Bellini:2014uqa}\cite{Keefer:2011kf}. This is due to the fact that the crude oil used in the production of the solvents and fluors has spent about 200\,Myrs in a subterranean reservoir, well shielded from biological activity and cosmic radiation.

However, following through on this argument, the expected level of \ce{^{14}C} in the scintillator is even lower: with a half-life of $T_{1/2}=5,730$\,yrs, an initial natural abundance of $R\sim10^{-12}$ and a storage time of hundreds of millions of years, the initial \ce{^{14}C} content will have decayed long ago. Consequently, the observed level must result from intermediate replenishment. In \cite{Alimonti:1998rc,Resconi:2001bxa}, the authors identify natural radioactivity from the rock surrounding the oil reservoir as the most likely source. A low but steady flux of neutrons is created by natural radioactivity in the rock, $(\alpha,n)$ reactions and spontaneous fission of \ce{^{238}U} being the main contributors. The neutrons undergo in turn $(n,p)$ reactions on \ce{^{14}N} nuclei trapped with the natural gas in the oil reservoir. The observed $R\sim$10$^{-18}$ are consistent with a (relatively high) \ce{^{14}N} content of $\sim$5\% (range: 0.2\% to 5.5\%)\footnote{https://www.uniongas.com} \cite{Resconi:2001bxa}. 

A simulation study performed in \cite{Resconi:2001bxa} suggests that the exact value of $R$ depends on the composition of the rocks surrounding the deposits and the amount of nitrogen present (that may locally depend on the depth inside the reservoir \cite{https://doi.org/10.48550/arxiv.hep-ex/0308025}). 
Following this line of thought, ideal conditions will be offered by an oil reservoir surrounded by low-radioactivity limestone (i.e.~low number of neutrons produced) that contains natural gas with a relatively low content of \ce{^{14}N} (i.e.~low number of targets for $(n,p)$-reactions). As a consequence, the extracted crude oil might feature an $R$ as low as $10^{-20}$ \cite{Resconi:2001bxa}. But even abundance levels of $5\cdot10^{-21}$~\cite{Alimonti:1998rc} or lower \cite{https://doi.org/10.48550/arxiv.hep-ex/0308025} seem plausible. It has been speculated that the higher $R$ values observed in neutrino experiments might be caused by contact to surface CO$_{2}$ during LS production or a \ce{^{14}C} contamination of the fluor that is added in small quantities to the solvent \cite{Alimonti:1998rc}, or might be caused by the biological activity of extremophilic bacteria \cite{https://doi.org/10.48550/arxiv.hep-ex/0308025}.


While these issues of finding low-\ce{^{14}C} crude oil have been discussed for several decades, they have never been put to rigorous systematic experimental testing beyond the $10^{-18}$ level. There are two generic problems to overcome: the identifying a suitable low-\ce{^{14}C} oil reservoir or provider and the manufacturing of a special batch of solvent (and fluor) for the full-scale experiment.

Measuring $R$ for a given LS sample at these low levels is difficult. In Ref.~\cite{Buck:2012zz}, the authors report measurements in a small but well-shielded LS chamber in an underground laboratory setup to observe the {\it in-situ} \ce{^{14}C} decays. This setup was able to report results with a sensitivity corresponding to $R\sim10^{-17}$. An experimental setup of similar dimensions but with an improved shielding concept is currently under preparation in the Pyh\"asalmi mine, Finland~\cite{Enqvist:2017xsa}. The aimed sensitivity of the Pyhäsalmi setup is between $R\sim10^{-17}$ and $10^{-18}$. To go beyond that requires a detector the size of Borexino's CTF or JUNO's OSIRIS \cite{Derbin:2004ma,JUNO:2021wzm} paired with the appropriate infrastructure for LS handling. Low-\ce{^{14}C} LS identification could  proceed in several steps: by studying the geology of oil fields, a small (liter-scale) batch of crude oil can be retrieved directly from that field and refined for use in the Pyh\"asalmi setup -- note that it is not necessary to produce a scintillator solvent since also alcanes mixed with a fluor will provide sufficient light yield \cite{wurm:2005}. Once a sample with \ce{^{14}C} below the sensitivity threshold of the setup has been identified, a larger sample on the scale of several tons can be prepared to do the ultimate test of $R$ on the scale of OSIRIS (i.e.~even before an upgrade to the Serappis setup of Sec.~\ref{sec:setup}).

It should be noted that standard practice of LS solvent producers is to intermix crude oil products from many different sources, i.e.,~oil reservoirs in fields. Production of a special batch of LS from only one particular low-\ce{^{14}C} oil field or reservoir will such mean a special order that will result in a substantial increase in production cost. While it is conceivable that this is financially possible for a detector the size of OSIRIS, it is for instance practically excluded for a full-scale neutrino detector like JUNO. The combination of the Pyh\"asalmi setup and OSIRIS such offers a comparatively cost-efficient way to first identify suitable low-\ce{^{14}C} solvent/fluor providers before conducting an experiment with a sufficiently small target mass to keep LS production costs under control.


\section{Simulation Study}
\label{sec:sim}

In order to predict signal and background levels of the Serappis experiment, we have performed a series of simulations. In Sec.~\ref{sec:internal}, we discuss the event rates and visible energy spectra to be observed for the recoil electrons from solar neutrinos and from radioactive decays of internal backgrounds, i.e.,~isotopes of the natural U/Th chains dissolved in the scintillator. Section ~\ref{sec:external} investigates the background created by gamma rays from radioactive decays outside the LS volume, mainly from the PMT glass and the cavern rock surrounding the detector setup. The results form the basis for the $pp$ neutrino sensitivity study presented in Sec.~\ref{sec:sensitivity}.

\subsection{Internal Events}
\label{sec:internal}


{\it Internal} events include both the electron recoils from solar neutrinos and radioactive decays, i.e.,~\ce{^{14}C} intrinsic to the LS molecules and U/Th chain isotopes dissolved in the liquid. For solar neutrinos, we regard only the $pp$ and \ce{^{7}Be} components. All other solar neutrino fluxes (CNO, $pep$, $^8$B) can be neglected in good approximation since there spectra extend to much higher energies and contribute event rates about two orders of magnitude smaller in the $pp$ signal region. Similarly to the situation observed in Borexino, we assume that the U/Th contamination of the final scintillator is sufficiently low (i.e.~on the level of $10^{-17}$\,g/g or less) so that the isotopes in secular equilibrium can be neglected in the spectral fit \cite{Bellini:2014uqa}. However, Borexino observed that secular equilibrium in the \ce{^{238}U} chain is broken for the long-lived \ce{^{210}Pb}, as well as its daughter nuclei \ce{^{210}Bi} and \ce{^{210}Po} \cite{Bellini:2014uqa}. All three are present at significantly higher activity levels as the basic U/Th concentration would suggest. While the $\beta$-decays of \ce{^{210}Pb} are too low in energy to be of interest for this analysis ($<$63\,keV), \ce{^{210}Bi} features a $Q$-value of 1.2\,MeV and thus contributes a virtually flat $\beta$-like background in the $pp$ energy range. By leaving it as a free component in the spectral fit (Sec.~\ref{sec:sensitivity}), the \ce{^{210}Bi} contribution is as well representative of other $\beta$-emitters that might be dissolved at low quantities in the LS, e.g. \ce{^{85}Kr}. The $\alpha$-emitter \ce{^{210}Po} features quite prominently in the Borexino low-energy spectrum but can be expected to have negligible impact given the envisaged energy resolution since there is no spectral overlap with the $pp$ neutrino flux. 

Moreover, we largely omit the decays of cosmogenic radioisotopes produced by muon spallation in the LS. At the relatively low overburden (1800\,mwe) of the JUNO underground laboratory, production rates will be comparatively high. However, all isotopes produced at high rates feature considerably higher energies than the $pp$ neutrinos and can be reduced by the sophisticated vetoing techniques developed for Borexino \cite{Bellini:2014uqa,Borexino:2021pyz}. Only the long-lived \ce{^{11}C} will likely give a visible rate contribution to the fit region. \ce{^{11}C} undergoes a $\beta^+$ decay corresponding to visible energies of 1--2\,MeV if both annihilation gamma-rays are observed (cf.~\cite{Bellini:2014uqa}). However, if one of the $\gamma$'s escapes into the water buffer, visible energy can become as low as 0.5\,MeV. So similarly to \ce{^{210}Po}, \ce{^{11}C} decays contribute to the signal region but do not influence the fit result since there is no direct overlap with the $pp$-spectrum.

To obtain the rates of the relevant spectral components, we scale the signal and background event rates extracted in the last Borexino $pp$-chain analysis \cite{Agostini:2018uly} to a fiducial LS volume corresponding to a 1.2\,m cylinder centered on the acrylic vessel, corresponding to 9.3 tons of LS (Sec.~\ref{sec:external}). We make an exception for \ce{^{14}C}, where we assume for our default scenario a significantly lower abundance of $10^{-20}$ that corresponds towards the lower expectation range for low-nitrogen oil reservoirs (Sec.~\ref{sec:c14}). The influence of all background rates on the $pp$ sensitivity are investigated in Sec.~\ref{sec:variation}.

\begin{figure}[t!]
\centering
\includegraphics[width=0.48\textwidth]{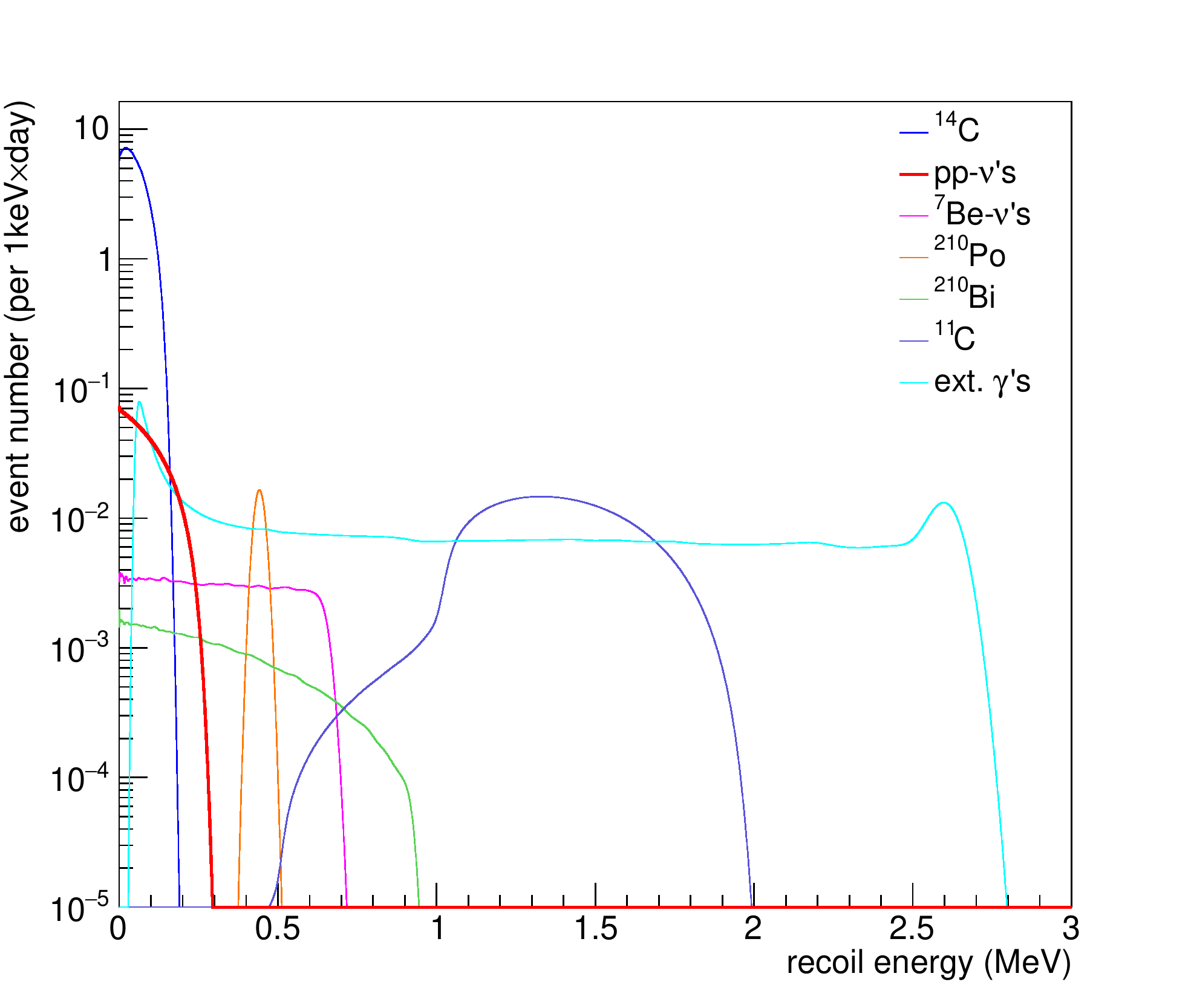}   
\caption{Energy spectra for all contributions explicitly regarded in the analysis, i.e.,~solar $pp$ and \ce{^{7}Be} neutrino events and \ce{^{14}C}, \ce{^{210}Po}, \ce{^{210}Bi} and \ce{^{11}C} background events. Rate normalizations correspond to a fiducial mass of 9.3 tons (Sec.~\ref{sec:external}). External backgrounds are not shown. The spectral shapes already take into account an energy resolution that corresponds to 1,500 p.e./MeV collected.
}
\label{fig:internal}
\end{figure}

The spectral fit presented in Sec.~\ref{sec:sensitivity} relies on the energy spectra of these components, determining their contributions based on the characteristic spectral features. The underlying visible energy spectra are shown in Fig.~\ref{fig:internal} and the corresponding total rates are summarized in Tab.~\ref{tab:wbls}. They are derived from a simplified Monte-Carlo simulation that includes both light emission and propagation in the LS volume and detection by the PMTs. According to the layout described for Serappis in Sec.~\ref{sec:serappis}, the model considers 74 20-inch PMTs equipped with light-collecting cones with opening diameters of 1\,m, located 2.5\,m from the detector center. The scintillation light yield $Y_{ph}$ is assumed to be $10^4$ photons/MeV for $\beta$-like events. The number of photons created per event is taken from a Gaussian distribution with $\sigma = \sqrt{E\cdot Y_{ph}}$. Photons are emitted isotropically. Absorption and scattering losses in the detection media are neglected because of the small detector dimensions. When falling onto a cone opening, we assume a detection efficiency of 30\%, corresponding to the quantum efficiency of the OSIRIS PMTs and neglecting reflection losses on the cones. 

To estimate the expected energy resolution of the setup, we use the average number of photoelectrons detected for events isotropically distributed within the LS volume: in the described optimum configuration, we obtain a photo electron yield $Y_{pe}$ of slightly more than  2,000\,p.e./MeV. Conservatively, we assume $Y_{pe}=1,500$\ p.e./MeV in our default model (cf. Fig.~\ref{fig:internal}). In addition, we include the effect of the limited charge resolution of the PMTs, that translates to an additional stochastic contribution corresponding to 14\% of the photon statistics. Despite the low energy range, the contribution from dark noise ($\sim 1.8\cdot10^4\,{\rm s}^{-1}$ per PMT) can be safely neglected because of the small number of PMTs. For the \ce{^{210}Po} $\alpha$-decays, we assume a (conservatively high) quenching factor of $\sim$12 but find no overlap with the $pp$ spectrum. No further systematic effects have been included. 

The vertex position was reconstructed using PMT time differences, including a transit time spread of 1.15\,ns ($1\sigma$) and the photon emission profile from Ref. \cite{OKeeffe:2011dex}. The precision of the reconstruction is in good agreement with the values found for Borexino for equivalent p.e.~numbers and PMTs with a very similar timing performance \cite{Back:2012awa}. In the following, we assume a Gaussian vertex resolution with a 1-dimensional $\sigma$ of 7\,cm at 1,000 p.e.~and scaling with the inverse of the square root of collected p.e. number.



\begin{table}[h!]
\centering
\caption{Signal and background rates for a cylindrical fiducial volume of 2.4\,m height and diameter (9.3\,t). Interaction rates measured in Borexino were used for $pp$ and \ce{^{7}Be} neutrinos as well as \ce{^{210}Bi} background \cite{Agostini:2018uly}. \ce{^{210}Po} is assumed in secular equilibrium but is irrelevant to the fit. The same is true for cosmogenic \ce{^{11}C}. The \ce{^{14}C} abundance assumed here is $10^{-20}$. For external background see Sec.~\ref{sec:external}}
\label{tab:wbls}
\begin{tabular}{lr}
\hline
Contribution & Rate (d$^{-1}$)\\
\hline
$pp$ neutrinos & 12.5 \\
\ce{^{7}Be} neutrinos & 4.5 \\
\ce{^{14}C} & 1,360 \\
\ce{^{210}Po} & 1.6 \\
\ce{^{210}Bi} & 1.6 \\
\ce{^{11}C} & 13.3 \\
external $\gamma$'s & 35 \\
\hline
\end{tabular}
\end{table}


\begin{table*}[h!]
    \centering 
    \caption{Limits on radioactive contamination of the OSIRIS detector materials, estimated background levels for OSIRIS and Serappis including the effect of the improved shielding (f.ex water, low-background concrete and/or steel). We assume that, in case the Acrylic Vessel turns out to be close to OSIRIS specification limits, it will be replaced for Serappis.}
    \label{tab:external_bg}
    \begin{tabular}{l|rrrr|rr}
                    & \multicolumn{3}{c}{Contamination [Bq/kg]} &  &  \multicolumn{2}{c}{Rate in LS [s$^{-1}$]}\\
        Component   & \ce{^{40}K}  & \ce{^{232}Th}    & \ce{^{238}U} & Mass & OSIRIS & Serappis \\
        \hline
        Cavern Rock & $2.2\cdot10^2$ &  $1.23\cdot10^2$ &  $1.42\cdot10^2$  & $-$ & 3.8 & $2\cdot10^{-3}$\\
        Water Tank & $2.6\cdot10^{-1}$ &  $4\cdot10^{-2}$ &  $7\cdot10^{-2}$ & 36\,t &  $1.2\cdot10^{-4}$ & $1.2\cdot10^{-4}$\\
        PMT Glass   & 1.9 & 1.7 & 4.8 & 0.6\,t & 1.5 &  $0.8\cdot10^{-3}$ \\
        Steel Frame & 0.27 & 0.16 & 0.24 & 3\,t &  $2.4\cdot10^{-2}$ & $1.2\cdot10^{-5}$\\
        Acrylic Vessel & $5.3\cdot10^{-4}$ &  $9.1\cdot10^{-5}$ &  $1.1\cdot10^{-6}$ & 1.3\,t &  $2.3\cdot10^{-2}$ 
        & negligible 
        \\
    \end{tabular}  
\end{table*}

\subsection{External gamma-ray background}
\label{sec:external}

Radioactive decays in the outer detector materials and in the rock surrounding the detector are inducing a gamma background to neutrino detection. To estimate this {\it external} background level, we consider all isotopes of the $^{238}$U and $^{232}$Th decay chains assuming secular equilibrium, as well as $^{40}$K. The isotopic abundance levels are assumed according to the design requirements of the OSIRIS detector \cite{Genster:2019ier}. 

We estimate the background rate in Serappis using a GEANT4-based study of the external gamma-ray event rate that has been performed for OSIRIS \cite{Genster:2019ier}. For the inner materials, i.e.~the Acrylic Vessel, water, PMT glass, and Steel Frame, it is in principle possible to use a "brute-force" simulation, in which gamma rays are generated in the respective starting media and then propagated until the interaction in the target LS. The detector design relies on extensive shielding by the water buffer. Attenuation factors are as large as $\sim10^{10}$ when it comes to gammas emitted by the Steel Tank and the surrounding rock, preventing the accumulation of meaningful event statistics by the standard method. Instead, we have implemented a geometrical biasing approach that is provided as part of the GEANT4 framework \cite{Genster:2019ier,Allison:2016lfl}. For events from the rock, our implementation of the biasing approach offers an efficiency gain of over $10^6$ compared to the ''standard'' simulation.

The corresponding external background rates in OSIRIS are listed in Table ~\ref{tab:external_bg}. In standard OSIRIS dimensions, the dominating external gamma signal rate is caused by the cavern rock, amounting to 3.8 background counts per second in the LS volume. The PMTs have been placed sufficiently far from the LS volume to contribute only 1.5 events per second. All other contributions are negligible. While these rates are sufficiently low for the measurement program of OSIRIS, overcoming these background levels will make a $pp$ neutrino detection very challenging. This is a main driver for the changes in geometric design that are lined out in Sec.~\ref{sec:serappis}. By adding shielding around the water tank and removing the PMTs further from the LS volume, background rates can be reduced to a level of 2.8\,mBq (Tab.~\ref{tab:external_bg}). This rate has been obtained by scaling the contributions of the main gamma branches regarded in the OSIRIS biasing study by the gamma attenuation in 1.2\,m of water expected at the respective energies. A further issue is the Acrylic Vessel. Based on the radiopurity levels that have been measured during its production (Tab.~\ref{tab:external_bg}), the rather sturdy vessel with 3\,cm of wall thickness will remain as the largest gamma background source in Serappis. Hence, a replacement (possibly by a thin-walled spherical vessel or nylon balloon) will have to be considered. We assume in the following that such a replacement is technically feasible and the background contribution of the new vessel will again be negligible compared to other external sources.


The output of the OSIRIS background simulation are true energy depositions in the LS volume. At these energies, gamma rays deposit their energy mostly based on Compton scattering. In agreement with standard MC treatment, we regard the barycenter of energy depositions as the true (i.e.~mean reconstructed) event vertex. It should be noted that gamma rays interacting close to the boundary of the LS volume may deposit only part of their energy inside the LS, leading to a reduction in visible energy. This is a very relevant effect for $pp$ neutrino detection since these events populate the  low-energy region of the visible energy spectrum. However, vertices are expected to be reconstructed close to the vessel boundary and can be removed efficiently by a fiducial volume cut.



\begin{figure*}[h]
    \centering
    \includegraphics[width=\textwidth]{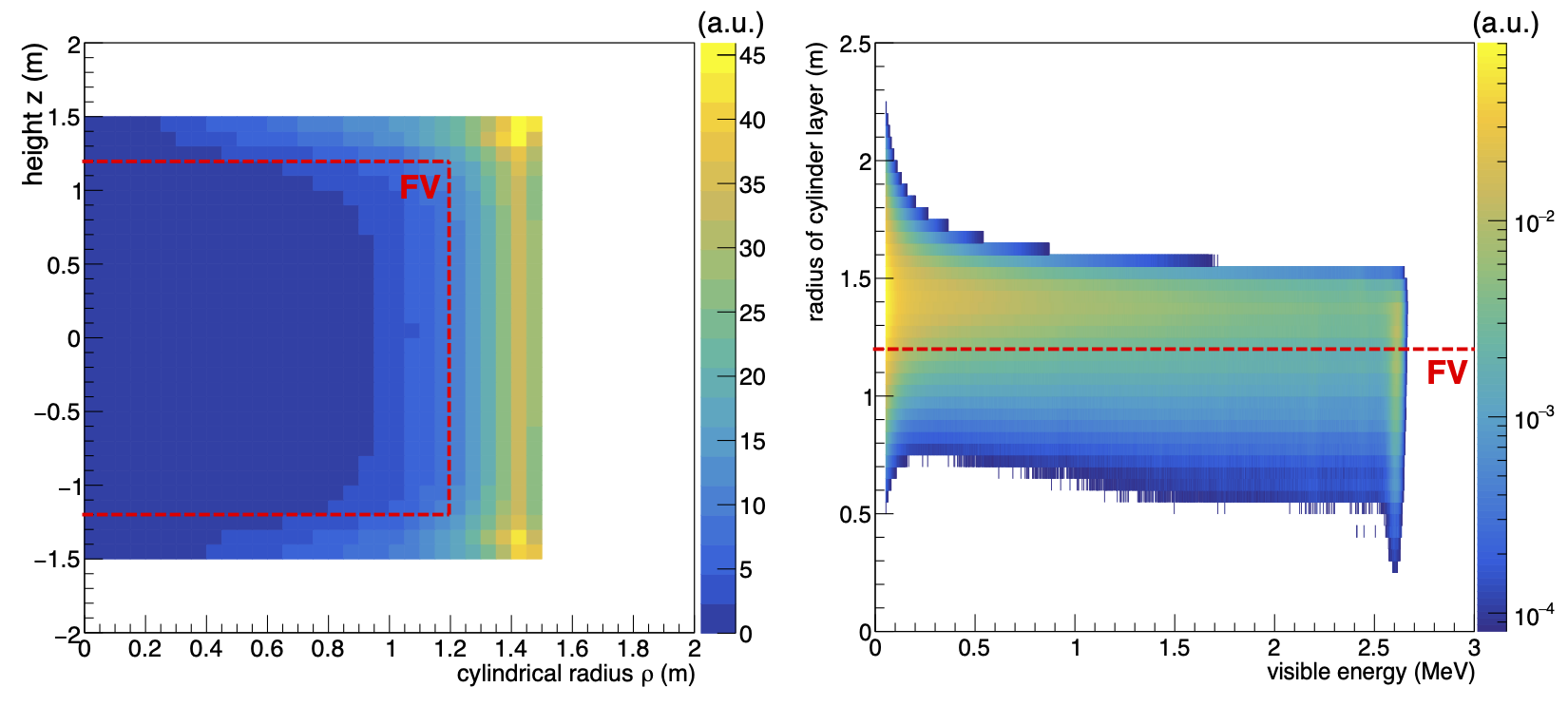}
    \caption{Event distributions of the external gamma-ray background, assuming a photoelectron yield of 1,500 pe/MeV and Gaussian energy and vertex smearing. {\it Left:} spatial distribution of true energy depositions inside the Acrylic Vessel. {\it Right:} cylinder-layer distribution of reconstructed vertices as a function of reconstructed gamma energy. Note the enhanced radial smearing of vertices at the lowest energies. The default fiducial volume is indicated with the red dashed line}
    \label{fig:external_gammas}
\end{figure*}

To include this effect, we smear the reconstructed position of the $\gamma$ vertices according to the simplified treatment described in Sec.~\ref{sec:internal}, taking into account the visible energy deposition. The resulting event distributions inside the LS volume are displayed in Fig.~\ref{fig:external_gammas}: the left panel displays a two-dimensional projection of the vertex positions on the cylinder axis and radius, the right panel shows the reconstructed energy distribution for the events assorted by cylinder layers. Obviously, most of the gamma events are expected to be reconstructed close to the verge of the LS vessel, facilitating a significant reduction of the external background rate based on a cylindrical fiducial volume cut. Based on these distributions and an optimization study for the fit uncertainties (Sec.~\ref{sec:sensitivity}), we apply a cut at 1.2\,m in cylinder radius and half-height to define the fiducial volume for the further analysis. At high energies, the 2.6\,MeV gammas emitted in the decay of \ce{^{208}Tl} are penetrating the deepest into the LS volume. On the other hand, event vertices are smeared out by the insufficient p.e.~statistics at the low-energy end of the spectrum. Both features are clearly visible in the external gamma spectrum shown in Fig.~\ref{fig:default_fit}. 
 

\section{Expected Sensitivity}
\label{sec:sensitivity}

\begin{figure*}[htb!]
    \centering
    \includegraphics[width=0.75\textwidth]{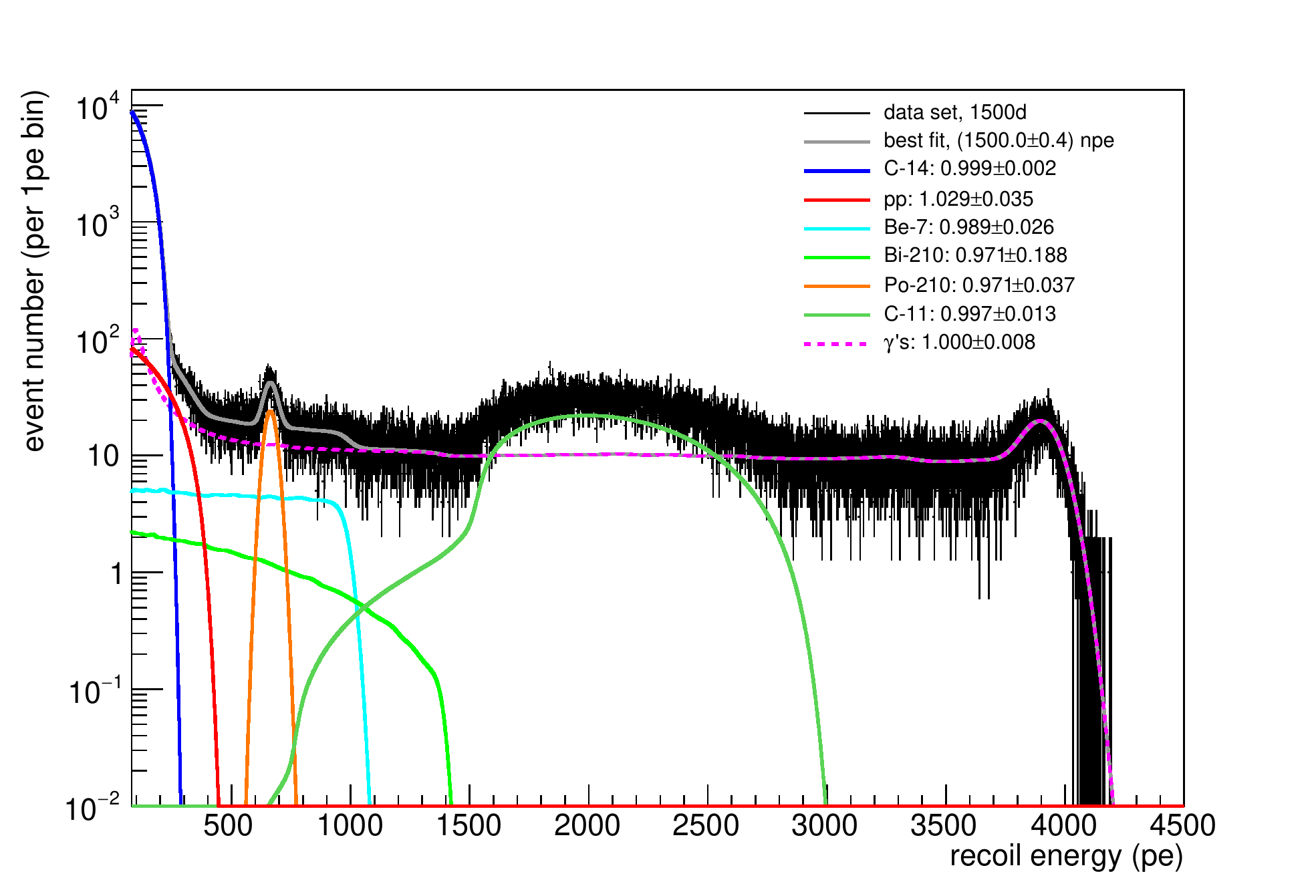}
    \caption{Energy spectrum and spectral fit as expected for the default scenario of Serappis (see text). Both the data spectrum and the fit include only the spectral components potentially relevant for the fit uncertainty of the $pp$ rate, i.e.~$pp$ and \ce{^{7}Be} recoil spectra as well as the dominant internal radioactive backgrounds \ce{^{14}C}, \ce{^{210}Bi}, \ce{^{210}Po}, cosmogenic \ce{^{11}C} as well as external $\gamma$ rays. The resulting relative uncertainty for the $pp$ rate is 3.4\,\% for a 1500-day measuring time. Crucially, pile-up events from \ce{^{14}C} that are of special importance for Borexino \cite{Bellini:2014uqa} can be neglected for Serappis due to the overall low event rates.}
    \label{fig:default_fit}
\end{figure*}

To determine the potential sensitivity of Serappis to the $pp$ neutrino flux, we perform  spectral fits to MC data sets that are largely inspired by the corresponding Borexino $pp$ analysis. In our somewhat simplified fit model, we take into account the seven most relevant low-energy neutrino signals and radioactive backgrounds inside the fiducial volume (Figure \ref{fig:default_fit}): solar $pp$ and \ce{^{7}Be} neutrinos, \ce{^{210}Bi}, \ce{^{210}Po}, \ce{^{11}C}, external gammas and of course \ce{^{14}C}. As laid out in section \ref{sec:internal}, we do not include all other solar neutrino species since their contributions in the low-energy spectral regions are negligible. Moreover, for the purpose of the $pp$ fit a spectral contribution by \ce{^{85}Kr} is largely exchangeable with a higher \ce{^{210}Bi} background and is thus not considered separately. For \ce{^{210}Bi}, we assumed a decay rate of 0.18 per day and ton, corresponding to Borexino levels \cite{Agostini:2018uly}. \ce{^{210}Po} is included at the same rate, \ce{^{11}C} at the rate expected for the overburden of the JUNO site. However, since there is no spectral overlap of \ce{^{210}Po} and \ce{^{11}C}  with the $pp$ recoil electrons, they do not influence the fit result.  

Finally, we do not include the background from \ce{^{14}C} pile up since at the lower \ce{^{14}C} rates and volume of Serappis, the corresponding rate can be expected to be many orders of magnitude smaller than in Borexino where \ce{^{14}C} pile up accounts for about 10\% of the events in the spectral gap between \ce{^{14}C} and \ce{^{210}Po} spectra \cite{Bellini:2014uqa}. Any reduction in absolute \ce{^{14}C} rate enters the rate of accidental \ce{^{14}C}--\ce{^{14}C} coincidences squared, so that the ten-fold smaller fiducial mass of Serappis alone means a rate reduction by a factor $\sim$10$^2$ compared to Borexino. This would be further reduced by a value of $R(\ce{^{14}C})$ below that of Borexino. Hence, we conclude that \ce{^{14}C} pile up is not relevant for the spectral fit.

Under these assumptions, we perform a log-likelihood fit of the seven components $j$ to the MC data spectrum: 
\begin{eqnarray}
-2\log{\cal L}(n_j) = \sum_{i} -2\log P(n_i,\mu_i) + \sum_j \left( \frac{(n_j-1)\hat\mu_{j}}{\sigma_j} \right)^2
\label{eq:fit}
\end{eqnarray}
 The free fit parameters are the normalization factors $n_j$ relative to the input rates in the data spectrum. The first term denotes the fit to the data values in the energy bins: the expectation value of the fit in the {\it i-th} energy bin is denoted by $\mu_i = \sum_j n_j\mu_{i,j}$, with $\mu_{i,j}$ the spectral contribution of a specific fit component $j$. The second sum represents optional penalty terms that are based on independently determined background rates $\hat\mu_j$ with uncertainties $\sigma_j$ (see below). As described in Sec.~\ref{sec:sim}, the spectra used for the fit components are derived as well from MC simulations.

Figure \ref{fig:default_fit} shows an exemplary fit to a data spectrum including statistical variations. It uses a lower energy limit of 50\,keV to avoid threshold effects. Comparable to the situation in Borexino, \ce{^{14}C} is dominating in the low-energy regime, while \ce{^{7}Be} neutrinos and external gammas are the largest contributions at higher energies. The $pp$ neutrino signal emerges from background only in a narrow energy region at the foot of the \ce{^{14}C} spectrum. Note that the sensitivity studies described below mostly use the Asimov data set to obtain a median expectation value. The validity of the simplification in this special case  has been tested by fitting 1000 statistically smeared data sets according to our default scenario (see below). 

\subsection{Default scenario} 
\label{sec:default}

Based on the experimental layout of Serappis presented in Section \ref{sec:setup}, we define a default scenario for the spectral fits. For this, we assume a liquid scintillator (LS) with a \ce{^{14}C} abundance of $R(\ce{^{14}C})=10^{-20}$. Light collection is enhanced by adding Winston cones resulting in a photoelectron yield of $Y_{pe}=1,500$\,p.e./MeV. The external shielding to the cylindrical LS volume is 4.2\,m of water equivalent. The corresponding optimum fiducial volume is $\rho,|z|<120$ cm, i.e.~forfeiting the outer 30\,cm of the scintillator volume to reduce the background from external $\gamma$-rays (and from possible contaminations on the vessel surface) to an acceptable level. The corresponding fiducial mass is 9.3 tons. 

To obtain the sensitivity of this default scenario, some further choices have to be made for the fit.  For the \ce{^{14}C} and external gamma rates are left free in the fit,
while the \ce{^{7}Be} rate measurement by Borexino is included as a constraint. As total measurement time we choose $T=1,500$\,days, corresponding roughly to a 5-years measurement. The exact value of the p.e.~yield $Y_{pe}$ is left as a free parameter in the fit and self-calibrates based on the \ce{^{14}C} spectrum and the \ce{^{208}Tl} peak. Similarly to the situation in Borexino, the fit threshold is set to 50\,keV \cite{Bellini:2014uqa}. In this default configuration, Serappis is able to measure the $pp$ interaction rate with a relative uncertainty of $\delta_{pp}=3.4\%$, i.e.~reaching into the range that will be interesting from the point of view of neutrino oscillations.

\subsection{Dependence on detector performance} 
\label{sec:variation}

\noindent To explore the relative importance of the detector performance on the final sensitivity, we extended the study by varying the corresponding parameters within their conceivable bounds. The results for $\delta_{pp}$ are summarized in Table \ref{tab:delta_pp}.

\begin{itemize}
\item[\bf (a)]{\bf \ce{^{14}C}/\ce{^{12}C} abundance ratio $R$.} Key to the Serappis approach is the use of a low-\ce{^{14}C} LS for the neutrino target. We assume $R\sim10^{-20}$ as a default that corresponds to the lower end given in Ref.~\cite{Resconi:2001bxa}. However, even lower values of $R$ might be achieved in case of a reservoir with extremely low nitrogen content or surrounding radioactivity. Hence, we vary $R$ in a range from $10^{-22}$ to the $10^{-18}$ level achieved in the Borexino CTF. The corresponding range in $\delta_{pp}$ is 2.5\% to 5.0\%.\\
It is worth to note that even at $R\sim10^{-18}$, the sensitivity of Serappis would be about a factor 2 better than Borexino's. Pushing the measuring time and p.e.~yield to their optimum values results in a sensitivity of $\delta_{pp}\sim3.3\%$. A precision $pp$ measurement is thus possible even without identifying a low-\ce{^{14}C} scintillator but limited to the 3\% level.

\item[\bf (b)]{\bf Photoelectron yield $Y_{pe}$.} High energy resolution reduces the overlap between the \ce{^{14}C} and $pp$ neutrino spectra and thus facilitates the separation of the two components by the fit. We scan a range from 250\,pe/MeV, corresponding to the yield expected for the OSIRIS setup, to 2,000\,pe MeV which assumes $\sim$67\% optical coverage. While there is a substantial increase in $\delta_{pp}$ sensitivity from 6\% to 4\% in the range below 1,000\,pe/MeV, sensitivity almost stabilizes in the range above. Therefore, the default $Y_{pe}=1500$\, pe/MeV seems to reside close to the sweet spot between experimental effort and gain in sensitivity.

\item[\bf (c)]{\bf Measuring time $T$.} The accumulation of signal statistics has a great effect on $pp$ sensitivity. While Serappis profits from the relative smallness in many aspects (LS preparation, light collection, absence of \ce{^{14}C} pile-up), sensitivity is too a large part dependent on the collected statistics. We varied $T$ in steps of 500\,d up to a maximum duration of 3,000 days, corresponding to maybe 10 years of detector operation and $\delta_{pp}\sim2.5\%$. Experience gained in Borexino shows that it is very hard to keep the detector performance stable over such a long period of time -- the largest data sets analyzed are about 5 years, corresponding to the default value for $T$ assumed here.

\item[\bf (d)]{\bf Internal \ce{^{210}Bi} backgrounds.} A further important factor for the final sensitivity is the internal background. As explained above, this mostly concerns the levels of $\beta$-decays of \ce{^{210}Bi} and \ce{^{85}Kr} that are somewhat interchangeable. We vary the corresponding rate in the range from $10^{-2}$ to 100 decays per ton LS and day, which corresponds roughly to an order of magnitude better and two orders worse than the situation in Borexino. While lower \ce{^{210}Bi} levels have almost no effect, the $pp$ sensitivity starts to deteriorate for background rates above 1 decay per day and ton.\\
Clearly, achieving radiopurity levels comparable to the ones in Borexino is not a small feat. However, Serappis will be able to profit from the extensive infrastructure available for the purification of the JUNO scintillator. What is more, the amount of LS required is quite low, enabling repeated purification of the entire target volume while monitoring the progress made on \ce{^{210}Bi} levels. A \ce{^{210}Bi} on the level of Borexino can thus be considered to be within reach.

\item[\bf (e)]{\bf External constraints on \ce{^{14}C} and gamma background.} In most of the scenarios investigated above, the application of penalties on \ce{^{14}C} and gamma normalization as part of the likelihood (\ref{eq:fit}) has only a minor impact on the result. However, the situation changes considerably once the basic sensitivity of the scenario reaches the 2\% mark. At this stage, putting external constraints can decisively improve sensitivity. This situation is investigated in the 5th row (e) of Tab.~\ref{tab:delta_pp}: Starting from a very optimistic scenario of ultra-low \ce{^{14}C} abundance $R\sim10^{-22}$, long measuring time $T=3,000$\,d and optimum light yield $Y_{pe}=2,000$\ pe/MeV, $\delta_{pp}$ reaches an accuracy of 1.9\%. Adding a Borexino-like prior of 2\% on both \ce{^{14}C} and gamma rate will improve this result to 1.5\%. For 0.1\% priors, accuracy would reach 1.2\%.\\
As described in Ref.~\cite{Bellini:2014uqa}, the $pp$ analysis in Borexino uses a sample of untriggered events of \ce{^{14}C} decays to obtain a 2.2\% prior on the decay rate. Some improvement seems likely to be feasible, although a result beyond 1\% seems very hard to achieve. Similarly, the gamma background rate can be calibrated to high precision by studying the event distribution in the outer layers of the LS volume that are not part of the fiducial volume and extrapolating to the inside. This will be aided by the rigid target geometry of the Serappis acrylic vessel.
\end{itemize}

\begin{table*}[t!]
    \centering 
    \caption{Dependence of the relative $pp$ flux uncertainty measured by Serappis, $\delta_{pp}$, on (a) the \ce{^{14}C} content of the scintillator, (b) the photo electron yield, (c) the measuring time $T$, and (d) the level of internal \ce{^{210}Bi} background. All other parameters are kept according to the default scenario. For row (e), $R=10^{-22}$, $Y_{pe}=2,000$\,pe/MeV and $T=3,000$\,d is assumed while studying the impact of external constraints on the \ce{^{14}C} and $\gamma$ rates. Row (f) studies the dependence of the result on an external constraint on the Weinberg angle $\theta_W$. All values quoted are median sensitivities.}
    \label{tab:delta_pp}
    \begin{tabular}{rr|cccccc}
        {\bf (a)} & $R(\ce{^{14}C})$ & Borex & $10^{-18}$ & $10^{-19}$ & $\bm{10^{-20}}$ & $10^{-21}$ & $10^{-22}$ \\ 
        \cline{2-8}
        &  $\delta_{pp}$ & 5.6\% & 5.0\% & 4.2\% & {\bf 3.4\%} & 2.8\% & 2.5\%\\
        ~\\
        {\bf (b)} & $Y_{pe}$ [pe/MeV]& 250 & 500 & 750 & 1,000 & \bf 1,500 & 2,000 \\
         \cline{2-8}
        & $\delta_{pp}$ & 4.3\% & 4.1\% & 3.7\% & 3.6\% & {\bf 3.4\%} & 3.3\%\\
        ~\\
        {\bf (c)} & $T$ [d] & 500 & 1,000 & \bf 1,500 & 2,000 & 2,500 & 3,000 \\ 
         \cline{2-8}
        & $\delta_{pp}$ & 6.1\% & 4.2\% & {\bf 3.4\%} & 3.0\% & 2.7\% & 2.5\%\\
        ~\\
        {\bf (d)} & \ce{^{210}Bi} [/(dt)] & $10^{-2}$ & {\bf 0.16} & 0.5 & 1 & 5 & 10 \\ \cline{2-8}
        & $\delta_{pp}$ & 3.4\% & {\bf 3.4\%} & 3.6\% & 3.8\% & 4.9\% & 5.8\%\\
        ~\\
        {\bf (e)} &  $\sigma(\ce{^{14}C},\gamma)$ & \bf 1 & 0.1 & 0.05 & 0.02 &  0.01 & $10^{-3}$ \\ 
        \cline{2-8}
        & $\delta_{pp}$ &  1.9\% & 1.8\% & 1.7\% & 1.5\% & 1.3\% & 1.2\% \\
        ~\\
        {\bf (f)} &  $\sigma(\theta_W)$ & \bf 0.07\% & 0.1\% & 1\% &  5\%  & 10\% & 100\%\\ 
        \cline{2-8}
        & $\delta_{pp}$ &  \bf 3.4\% & 3.4\% & 3.5\% & 3.6\% & 3.8\% & 13\% \\
    \end{tabular}  
\end{table*}

In summary, the following conclusions can be drawn from the above discussion: if only one of the decisive experimental aspects (i.e.,~\ce{^{14}C} abundance, measuring time) are improved, sensitivity of Serappis could reach the 2.5\% level. If both and the energy resolution are improved, the $pp$ rate uncertainty can be reduced below the 2\% level. In this case, the application of external priors has the potential to reduce the effective uncertainty close to the 1\% level, i.e. comparable to SSM level uncertainties of the $pp$ flux prediction.

\subsection{Elastic neutrino-electron scattering and the Weinberg angle}
\label{sec:weinberg}

As lined out in Section \ref{sec:intro}, the low-energy differential cross-section for elastic scattering of neutrinos off electrons features a slight dependence on the exact value of the Weinberg angle $\theta_W$. In Ref.~\cite{Aalbers:2020gsn}, a study has been carried out for DARWIN to understand how well the $\nu_e$ survival probability and $\theta_W$ can be constrained by measuring the low-energy end of the electron recoil spectrum. It is found that $\sin^2\theta_W$ can be constrained on the 4\% level. While this is much less precise than theoretical predictions (e.g.~of $\sin^2\theta_W = 0.23867 \pm 0.00016$ from \cite{Erler:2004in}) it would still be an important measurement since no other experimental data on $\theta_W$ is available in this energy range.

To investigate the impact $\theta_W$ might have on a $pp$ measurement with Serappis, we have studied the dependence of the expected $pp$ interaction rate and spectral shape. For this, we rely on the $pp$ recoil spectrum including radiative corrections presented in Ref.~\cite{Bahcall:1995mm}, that as well quantifies the impact of $\sin^2\theta_W$ on the spectral shape.   

To quantify the size of a possible effect, we include a parametrization of the $\sin^2\theta_W$ dependence of the $pp$ recoil spectrum to the fit. A change in  $\sin^2\theta_W$ effectively translates to a tilt of the electron recoil spectrum at the level of 0.1\% for a 6.5\% change in $\sin^2\theta_W$ \cite{Bahcall:1995mm}. Given the shape of the $pp$ recoil spectrum, this effect is most prominent at the low-energy end of the $pp$ spectrum, i.e.~features maximum impact in the DARWIN measurement while being of minor impact for Serappis. 

We add a prior to the likelihood (Sec.~\ref{eq:fit}) in order to study the effect of constraining this parameter. Inserting the expected value and uncertainties from theory, we consistently reproduce the results presented in Sections~\ref{sec:default} and \ref{sec:variation}. In Tab.~\ref{tab:delta_pp}\,(f), we quantify how far the constraint on $\sin^2\theta_W$ can be weakened before it interferes with the precision of the $pp$ neutrino rate measurement. As expected, the effect on the precision of a $pp$ flux measurement is largely invisible until the uncertainty reaches the 5\% level. A very large uncertainty of 10\% would have to be assumed to seriously affect the accuracy of the flux measurement.


\section{Conclusions and Outlook}
\label{sec:conclusions}

The Serappis concept offers a very competitive approach for a precision measurement of the solar $pp$ neutrino flux. The decisive performance parameters are excellent energy and spatial resolution, low external gamma and internal \ce{^{210}Bi} (\ce{^{85}Kr}) background levels, and the use of an organic liquid scintillator selected for ultra-low \ce{^{14}C} abundance. If those conditions are met, a measurement of the $pp$ neutrino rate at the 3.5\% level becomes possible within 5\,yrs. Depending on the exact detector performance, even greater precision in the 1\%$-$2\% range might be achievable. This level of accuracy will be of interest not only for neutrino oscillations but also for solar physics and the search for invisible energy loss processes in the sun.

The Serappis detector can be realized by a moderate upgrade of the OSIRIS facility. It will greatly profit from the availability of the invaluable infrastructure created for the JUNO experiment, both what concerns the possibilities for the pre-screening of low-\ce{^{14}C} scintillator samples and the purification of the LS from U/Th chain elements, especially of long-lived \ce{^{210}Pb} and its daughters \ce{^{210}Bi} and \ce{^{210}Po} that have proven to be the most relevant low-energy backgrounds in Borexino \cite{Bellini:2014uqa}. The upgraded facility could serve more than one purpose: improved energy resolution and shielding will not only be of benefit for $pp$ neutrinos but as well be valuable to test loaded LS samples for a future $0\nu\beta\beta$ phase of JUNO \cite{Zhao:2016brs}.


%

\begin{acknowledgements}
This work was supported by the Deutsche For-schungsgemeinschaft (DFG), the Helmholtz Association, and the Cluster of Excellence PRISMA$^+$ in Germany, and the Ministry of Science and Higher Education of Russian Federation within the financing program of large scientific projects of the “Science” National Project (grant no. 075-15-2020-778).
\end{acknowledgements}

\twocolumn
\bibliographystyle{spphys}       
\bibliography{main}   


\end{document}